\documentstyle[11pt,preprint]{aastex}
\begin{document}
\title{LIGO gravitational wave detection, primordial black holes and the near-IR cosmic infrared background anisotropies}

\author{A. Kashlinsky\altaffilmark{1},}
\altaffiltext{1}{
Code 665, Observational Cosmology Lab, NASA Goddard Space Flight Center, 
Greenbelt, MD 20771 and
SSAI, Lanham, MD 20770; email: Alexander.Kashlinsky@nasa.gov} 
%\date{}
%\oddsidemargin .3in
%\evensidemargin .1in
%\footheight 12pt \footskip 30pt
%\topmargin 0.cm
%\textheight 21.5cm
%\textwidth 16.cm
%\renewcommand \baselinestretch{1.5}

% list of standard definitions of symbols, figure commands

\def\plotone#1{\centering \leavevmode
\epsfxsize=\columnwidth \epsfbox{#1}}

\def\wisk#1{\ifmmode{#1}\else{$#1$}\fi}

\def\wm2sr {Wm$^{-2}$sr$^{-1}$ }		%new definition
\def\nw2m4sr2 {nW$^2$m$^{-4}$sr$^{-2}$\ }		% new definition
\def\nwm2sr {nWm$^{-2}$sr$^{-1}$\ }		% new definition
\def\nw2m4sr {nW$^2$m$^{-4}$sr$^{-1}$\ }
\def\Ncut {$N_{\rm cut}$\ }
\def\lt     {\wisk{<}}
\def\gt     {\wisk{>}}
\def\le     {\wisk{_<\atop^=}}
\def\ge     {\wisk{_>\atop^=}}
\def\lsim   {\wisk{_<\atop^{\sim}}}
\def\gsim   {\wisk{_>\atop^{\sim}}}
\def\kms    {\wisk{{\rm ~km~s^{-1}}}}
\def\Lsun   {\wisk{{\rm L_\odot}}}
\def\Msun   {\wisk{{\rm M_\odot}}}
\def\um     { $\mu$m\ }
\def\sig    {\wisk{\sigma}}
\def\etal   {{\sl et~al.\ }}
\def\eg	    {{\it e.g.\ }}
\def\ie     {{\it i.e.\ }}
\def\bsl    {\wisk{\backslash}}
\def\by     {\wisk{\times}}
\def\cosec {\wisk{\rm cosec}}
\def\mic {\wisk{ \mu{\rm m }}}

\def\amin   {\wisk{^\prime\ }}
\def\asec   {\wisk{^{\prime\prime}\ }}
\def\cc     {\wisk{{\rm cm^{-3}\ }}}
\def\deg     {\wisk{^\circ}}
\def\ddeg   {\wisk{{\rlap.}^\circ}}
\def\damin  {\wisk{{\rlap.}^\prime}}
\def\dasec  {\wisk{{\rlap.}^{\prime\prime}}}
\def\approxeq{$\sim \over =$}
\def\abouteq{$\sim \over -$}
\def\percm{cm$^{-1}$}
\def\percmsq{cm$^{-2}$}
\def\percmcub{cm$^{-3}$}
\def\perhz{Hz$^{-1}$}
\def\perpc{$\rm pc^{-1}$}
\def\persec{s$^{-1}$}
\def\peryr{yr$^{-1}$}
\def\te{$\rm T_e$}
\def\tenup#1{10$^{#1}$}
\def\to{\wisk{\rightarrow}}
\def\thin{\thinspace}
\def\uk{$\rm \mu K$}
\def\p{\vskip 13pt}
%\maketitle

\begin{abstract}
LIGO's discovery of a gravitational wave from two merging black holes (BHs) of similar masses rekindled suggestions that primordial BHs (PBHs) make up the dark matter (DM). If so,  PBHs would add a Poissonian isocurvature density fluctuation component to the inflation-produced adiabatic density fluctuations. For LIGO's BH parameters, this extra component would dominate the small-scale power responsible for collapse of early  DM halos at $z\gsim 10$, where first luminous sources formed. We quantify the resultant increase in high-$z$ abundances of collapsed halos that are suitable for producing the first generation of stars and luminous sources. The significantly increased abundance of the early halos would naturally explain the observed source-subtracted near-IR cosmic infrared background (CIB) fluctuations, which cannot be accounted for by known galaxy populations. For LIGO's BH parameters this increase is such that the observed CIB fluctuation levels at 2 to 5 \um\ can be produced if only a tiny fraction of baryons in the collapsed DM halos forms luminous sources. Gas accretion onto these PBHs in collapsed halos, where first stars should also form, would straightforwardly account for the observed high coherence between the CIB and unresolved cosmic X-ray background in soft X-rays. We discuss modifications possibly required  in the processes of first star formation if LIGO-type BHs indeed make up the bulk or all of DM. The arguments are valid only if the PBHs make up all, or at least most, of DM, but at the same time the mechanism appears inevitable if DM is made of PBHs.

\end{abstract}

\section{Introduction}

LIGO's recent discovery of the gravitational wave (GW) from an inspiralling binary black hole (BH) system of essentially equal mass BHs ($\sim 30M_\odot$) at $z\sim0.1$\citep{Abbott:2016} has led to suggestion that all or at least a significant part of the dark matter (DM) is made up of primordial BHs (PBH) \citep{Bird:2016,Clesse:2016}. In particular, \cite{Bird:2016} argue that this PBH mass range is not ruled out by astronomical observations and the observed rate at $\sim$(a few) Gpc$^{-3}$yr$^{-1}$ \citep{Abbott:2016a} can be accounted for if DM PBHs are distributed in dense, low velocity-dispersion concentrations which escaped merging. There is abundant motivation for PBHs forming in the very early Universe \citep{Hawking:1971,Carr:1974}, e.g. during  phase transition at the QCD epoch when horizon mass is of the right magnitude \citep{Jedamzik:1997}; see a nice overview of possible mechanisms in \cite{Mack:2007}. If PBHs indeed constitute the bulk or all of DM, they would contribute an additional Poissonian component to the power spectrum of the mass distribution from adiabatic fluctuations from the earlier inflationary era \citep{Afshordi:2003}. If so, this component would dominate small scales leading to significant modification of the history of collapse (and possibly formation of the first sources), resulting in greater rates of cosmic infrared background (CIB) production at $z\gt10$.

CIB contains emissions produced over the entire history of the Universe including from sources inaccessible to direct telescopic studies \citep[see review by][]{Kashlinsky:2005} with early stars and BHs contributing to its near-IR ($\sim1-5\mu$m) fluctuation component \citep{Kashlinsky:2004,Cooray:2004}. In this context, \cite{Kashlinsky:2005,Kashlinsky:2007a} have identified from deep {\it Spitzer} data significant CIB fluctuations remaining on sub-degree scales after subtracting individual galaxies to faint levels. The measurement was later extended to degree scales \citep{Kashlinsky:2012} and confirmed in subsequent analyses of {\it Akari} and {\it Spitzer} data \citep{Matsumoto:2011,Cooray:2012}. It is now established that these fluctuations cannot arise from remaining known galaxy populations \citep{Kashlinsky:2005a,Helgason:2012a} and it has been suggested that they arise from new populations at early epochs \citep{Kashlinsky:2005a,Kashlinsky:2007b,Yue:2013a}. This proposition is currently a subject of debate \citep{Cooray:2012,Gong:2015}, although the CIB fluctuations on relevant scales appear to be uncorrelated with the diffuse light in the visible produced by sources down to AB mag $>28$ \citep{Kashlinsky:2007}\footnote{The intrahalo-light-model \citep{Cooray:2012} assigns the CIB fluctuations to light from galactic stars dispersed in mergers at low to intermediate $z$, but, at least in its present form, fails to account for numerous prior observational CIB data \citep[][]{Helgason:2014,Kashlinsky:2015a}; see e.g. Fig. 14 in \cite{Kashlinsky:2015a}.}. It was established that the CIB fluctuations are coherent with unresolved soft X-ray CXB \citep{Cappelluti:2013} at levels greater than expected from remaining known populations \citep{Helgason:2014} and that the measured coherence levels require much higher proportions of BHs among the CIB sources than in the known populations. At the same time, \cite{Helgason:2016} have argued that if early populations were to produce the measured CIB signal that would require higher than expected efficiencies of early star formation (cf. \cite{Kashlinsky:2015a}).

In this {\it Letter} we point out that if indeed the LIGO discovery is indicative of PBHs making up the DM, the extra Poissonian isocurvature component of the fluctuations would lead to much greater rates of collapse at early times, which would naturally produce the observed levels of the CIB fluctuations. We briefly revisit the required near-IR CIB energetics in Sec. \ref{sec:cib} and the effects of the extra power component from PBHs on the collapse of the first halos in Sec. \ref{sec:pbh}. We discuss the effects PBHs and the extra power may have on the formation of first populations in Sec. \ref{sec:discussion}. The discussion below adopts cosmology with $(h,\Omega_{\rm tot}, \Omega_{\rm CDM}, \Omega_{\rm bar}, \sigma_8)= (0.7,1,0.23, 0.05, 0.9)$.

\section{CIB anisotropies vs high-$z$ modeling}
\label{sec:cib}

The observed CIB fluctuations reflect several aspects of the sources producing them: 1) the measured shot-noise power characterizes the typical flux-magnitude of the sources producing the large-scale (clustering) power; the fact that the arcminute fluctuations arise at very low shot-noise levels means that the individual sources must  be very faint consistent with their high-$z$ origin \citep{Kashlinsky:2007b}.  2) The angular shape of the CIB power spectrum on arcminute scales reflects the epochs spanned by the sources; the shape from the {\it Spitzer} data is consistent with high-$z$ origin within the current errors and the upcoming  {\it Euclid} all-sky survey would  further probe the epochs and history of emissions much more accurately \citep{Kashlinsky:2015}. 3) Given the angular power template, the amplitude of the fluctuations spectrum at some fiducial scale (we adopt $5'$ below)  reflects the overall abundance of the sources with fluxes constrained by 1) and 2) via the corresponding mean CIB flux.

CIB integrated/bolometric flux levels depend on three efficiency parameters: 1) the efficiency of collapse of halos suitable for forming luminous sources, or the mass-fraction of the Universe in these halos, denoted $f_{\rm Halo}$, 2) the formation efficiency of conversion of baryons inside each halo into luminous sources, $f_*$, and 3) the radiation efficiency of converting the rest mass into radiation for the luminous  sources inside the collapsed halos, $\epsilon$.

We now briefly revisit the arguments in \citet[][Sec. 2 there]{Kashlinsky:2015a} for a  general set of efficiency requirements for sources at high $z$ to reproduce the observed CIB anisotropies between 2 and 5 \um\footnote{We limit the range to where there is overall consistency between the CIB fluctuation results from various instruments ({\it AKARI} and {\it Spitzer}) and analyses. The situation at 1 to 1.6 \um\ is contradictory as discussed in detail in \citet[][Sec. 2.1.2]{Kashlinsky:2015a}: conflicting levels have been measured by, on the one hand the mutually consistent (at the same depth) deep 2MASS \citep{Kashlinsky:2002,Odenwald:2003} and NICMOS \citep{Thompson:2007a,Thompson:2007} analysis,  and, on the other, the much shallower CIBER \citep{Zemcov:2014} data.}. The integrated CIB fluctuation at $5'$ between 2 and 5 \um from the  {\it AKARI} to {\it Spitzer} bands is $\delta F_{2-5\mic}(5')  \simeq 0.09\ {\rm nW\ m}^{-2}\ {\rm sr}^{-1}$; this arises as excess over that from known galaxies remaining in the data \citep{Helgason:2016}.  Populations at high $z$ are strongly biased, span a short period of cosmic time, and would be expected to produce CIB with relative fluctuation amplitude of $\Delta_{5'} \sim 10\%$ on arcminute scales, which  
would then require producing $F_{\rm CIB}(2\!-\!5\mic) = \delta F_{2-5\mic}(5')/\Delta_{5'} \sim 1$ nW m$^{-2}$ sr$^{-1}$ \citep{Kashlinsky:2007b} in the integrated flux at near-IR wavelengths $\sim(2-5)\mic$. The Lyman cutoff would cut the emissions below the observer wavelengths $\sim (1+z)/10\;\mic$.

Let us assume that a fraction $f_{\rm Halo}$ of all matter in the Universe collapses, in halos capable of producing luminous sources, at a given redshift converting on average a fraction $f_*$ of the halo baryons into luminous sources. %We call $f_*$ the formation efficiency. 
The bolometric diffuse flux produced by these populations, after they have converted their mass-energy into radiation with radiation efficiency $\epsilon$, is $F_{\rm tot} \simeq  f_{\rm Halo}f_*(\frac{c}{4\pi}\epsilon\rho_{\rm bar} c^2) z_{\rm eff}^{-1}\simeq 
9.1\times 10^5 \epsilon f_{\rm Halo}f_*z_{\rm eff}^{-1} \;\frac{\Omega_{\rm bar}h^2}{0.0227} \;\; {\rm nW\ m}^{-2}\ {\rm sr}^{-1}$ where $z_{\rm eff}\equiv 1/\langle(1+z)^{-1}\rangle$ is a suitably averaged effective redshift factor which accounts for the radiation energy density decreasing with expansion as $\propto (1+z)^{-4}$ vs. the matter density $\propto (1+z)^{-3}$. The overall fraction of Universe's baryons needed to explain the CIB is $f_{\rm Halo}f_*$ \citep[see Sec. 2.3.2 in][]{Kashlinsky:2015a}. For massive stars, which are fully convective and radiate close to the Eddington limit, $\epsilon \simeq 0.007$ for the H-burning phase of a few Myrs per star. Accreting BHs can reach electromagnetic radiation efficiencies $\epsilon=0.4$ for a maximally rotating Kerr BH. If the integrated CIB fluctuation approximates the bolometric flux produced by these sources, the fraction of baryons that on average go into the sources inside each halo, is:
\begin{equation}
f_* =  0.1\times \left(\frac{f_{\rm Halo}}{0.01}\right)^{-1}\; \left(\frac{\epsilon}{0.01}\right)^{-1}\;\left(\frac{z_{\rm eff}}{10}\right) \left(\frac{\Delta_{5'}}{0.1}\right)^{-1} \left[\frac{F_{\rm CIB}(2\!-\!5\mic)}{F_{\rm tot}}\right]^{-1}
\label{eq:fraction_bc}
\end{equation}
Thus in order to produce the measure CIB at $z>10$ with ``reasonable" formation efficiencies ($f_*<10\%$) one requires a large fraction of matter in collapsed halos capable of producing luminous sources (see next section).

\cite{Helgason:2016} discuss the requirements of high-$z$ sources to produce the observed CIB fluctuations within the conventional, if necessarily simplified, framework of gravitational clustering and spherical collapse of adiabatic $\Lambda$CDM fluctuations. They conclude that 1) first galaxies, if extrapolated to $z>8$ from known UV luminosity functions would produce much less CIB fluctuation power than observed \citep[cf.][]{Cooray:2012a,Yue:2013}, and 2) at still higher $z$ (first) stars would have to form inside the collapsed halos at substantial formation efficiencies (converting $f_* \ga 5\%$ of the available baryons in collapsing halos) and be very massive ($\sim 500M_\odot$) if they are to explain the observed CIB anisotropies. \cite{Kashlinsky:2015} reproduce the observed {\it Spitzer} signal with massive early stars forming at the mean formation efficiency $f_* \simeq$4\% out to $z=10$. 

The ``high-mean-formation-efficiency" difficulty can ultimately be traced to a relative paucity of high-$z$ collapsed halos - with the parameters appropriate for star formation - due to the limited amount of power on the relevant scales set by the adiabatic $\Lambda$CDM matter fluctuations, which arose from the period of inflation. The next section discusses how the abundances of the high-$z$ collapsed halos are dramatically increased if PBHs constitute the DM, and reduce - by large factors - the efficiencies required to produce the observed CIB anisotropies.

\section{PBHs, small scale mass fluctuation power and first object collapse}
\label{sec:pbh}

LIGO's GW150914 originated at $z=0.09$ from the merger of two BHs of essentially identical masses at $36^{+5}_{-4}$ and $29\pm4$ $M_\odot$ \citep{Abbott:2016}. This mass range lies within the horizon mass-scale at $\sim 0.01-0.1$ Gev where various mechanisms for generating PBHs in the very early Universe operate, such as discussed by e.g. \cite{Jedamzik:1997}. \cite{Bird:2016} discuss how the observed detection rate, inferred from the so far single published event, can be made consistent with that expected from the PBHs making up the DM such that their comoving mean mass density, assumed constant since their formation until at least their possible later evolution (discussed in Sec. \ref{sec:discussion} and references therein), is given by 
\begin{equation}
n_{\rm PBH}= \frac{1}{M_{\rm PBH}} \Omega_{\rm CDM}\frac{ 3H_0^2}{8\pi G} \simeq 10^9 \left(\frac{M_{\rm PBH}}{30M_\odot}\right)^{-1} \left(\frac{\Omega_{\rm CDM}h^2}{0.1}\right) {\rm Mpc}^{-3}
\label{eq:n_pbh}
\end{equation}
Below we will assume, for simplicity, that all PBHs have identical mass. The arguments that follow can be generalized to a PBH mass distribution, such as in e.g. \cite{Carr:1975,Choptuik:1993}, with $M_{\rm PBH}$ being the effective mass leading to the overall $n_{\rm PBH}$ comoving number density. We note that this mass range is allowed by, although close to, the limits from the MACHO microlensing surveys \citep{Alcock:2001}. \cite{Ricotti:2007,Ricotti:2008} have argued that, if PBHs are  of this mass range, accretion onto them may violate {\it COBE}/FIRAS constraints on the CMB black-body energy spectrum, but as \cite{Bird:2016} discuss, such arguments are model-dependent and subject to complex physics assumptions. \cite{Afshordi:2003} limit $M_{\rm PBH}<4\times 10^4M_\odot$ from Ly-$\alpha$ forest data. 
%The arguments below are valid only if the PBHs make up all, or at least most, of DM, but at the same time the mechanism appears inevitable if DM is made of PBHs.

\begin{figure}[h!]
\includegraphics[width=5in]{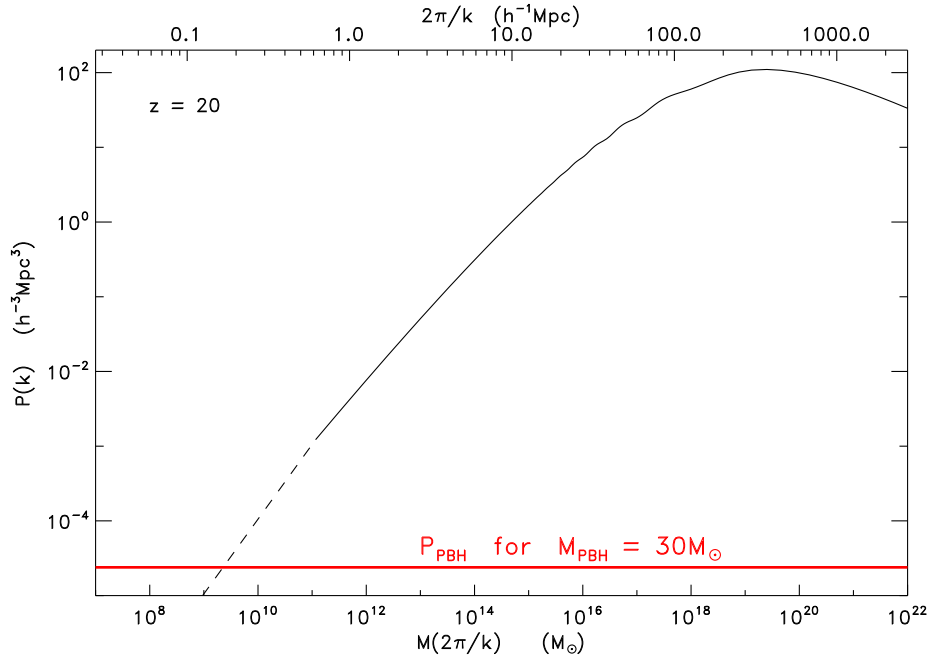}
 \caption[]{\footnotesize{Black solid line marks the CMBFAST-computed $\Lambda$CDM power spectrum at $z=20$ vs the mass contained within the comoving radius $2\pi/k$ for the cosmological parameters adopted here. Black dashes show the $P_{\Lambda{\rm CDM}}\propto k^{-3}$ extrapolation to scales inaccessible to CMBFAST, but relevant for the first halos collapse. Red horizontal solid line shows the Poissonian power from DM PBHs of $M_{\rm PBH}=30M_\odot$, which clearly dominates the scales relevant for halo collapse at this epoch.}}
\label{fig:fig1}
\end{figure}
As pointed out  by \cite{Afshordi:2003}, the DM from PBHs will contain an extra (isocurvature) component due to Poissonian fluctuations with the power component at the time of the PBH formation being $P_{\rm PBH, initial}=n_{PBH}^{-1}$  in comoving units. 
%This would result in an isocurvature component to the power. 
From their formation to today ($z=0$) these isocurvature fluctuations would grow, at wavelengths below the horizon at matter-radiation equality $z_{\rm eq}$, by a scale-independent factor of $\frac{3}{2}(1+z_{\rm eq})$, so the extra power component at redshift $z$  is given by \citep{Afshordi:2003}:
\begin{equation}
P_{\rm PBH}(z)=\frac{9}{4}(1+z_{\rm eq})^2n_{\rm PBH}^{-1} [g(z)]^{-2} \simeq 2\times10^{-2} \left(\frac{M_{\rm PBH}}{30M_\odot}\right) \left(\frac{\Omega_{\rm CDM}h^2}{0.13}\right)\left(\frac{1}{g^2(z)}\right)\;{\rm Mpc}^3
\label{eq:p_pbh}
\end{equation}
where $g(z)$ is the linear growth factor of fluctuations from $z$ to today, with $g(0)=1$. Fig. \ref{fig:fig1} shows the extra power component for $M_{\rm PBH}=30M_\odot$ compared to the $\Lambda$CDM power spectrum from the purely adiabatic fluctuation component. The power is plotted vs the mass contained in wavelength $2\pi/k$ which is $M(r) =1.15\times10^{12} (r/1{\rm Mpc})^3 M_\odot$ for the adopted cosmological parameters. This extra power is $\propto M_{\rm PBH}$ and for $M_{\rm PBH}>1M_\odot$ dominates the small scales relevant for collapse of the first halos at $z>10$. This  isocurvature power component dominates very small scales and has no impact on the observed CMB anisotropies or baryonic-acoustic-oscillations \citep[][]{Eisenstein:1999} which appear in CIB fluctuations on arcminute scales  and can be probed with Lyman-tomography of CIB from the upcoming {\it Euclid} survey \citep[][]{Kashlinsky:2015}. Furthermore, unlike the part of the power from clustering, white noise power contributions to the angular CIB power spectrum are not affected by biasing amplification \citep{Kashlinsky:2004}.

\begin{figure}[h!]
\includegraphics[width=5in]{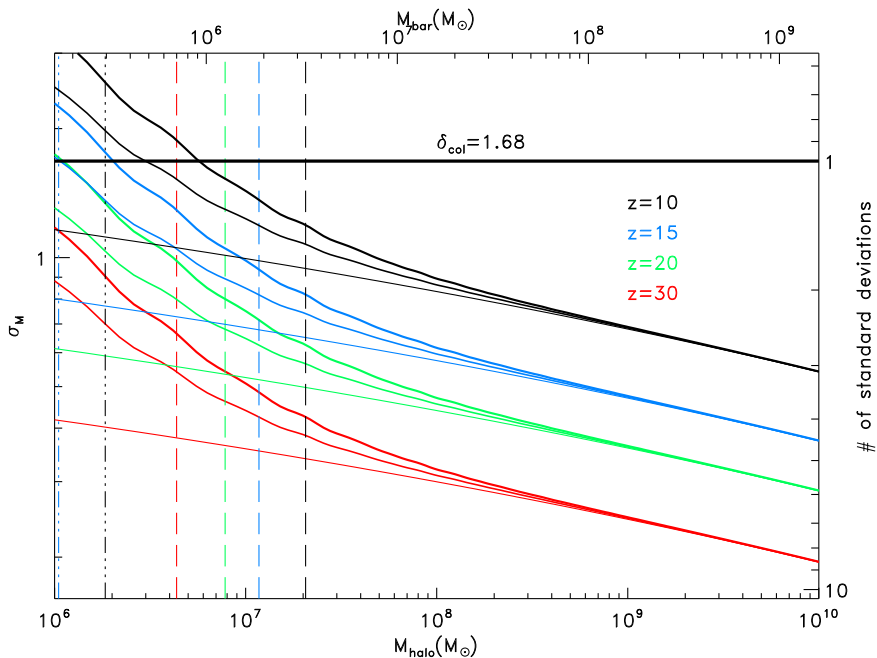}
 \caption[]{\footnotesize{Curves show the rms density contrast over the halo mass for $M_{\rm PBH}=0$ (thin), 15 (thick), 30 (thickest) $M_\odot$ at $z=$ 30 (red), 20 (green), 15 (blue), 10 (black). Black horizontal line shows $\delta_{\rm col}$, so halos with density contrast $>\delta_{\rm col}$ collapse at that $z$. Vertical dashes with same color notation mark halo mass where $T_{\rm vir}>10^4$K and vertical dash--dotted lines show the same for $T_{\rm vir}>10^3$K (at $z>15$ they are to the left of the box).}}
\label{fig:fig2}
\end{figure}
The net power spectrum would be given by $P_{\rm tot}(k,z) =P_{\Lambda{\rm CDM}} (k,z)+ P_{\rm PBH}(z)$, which we use to evaluate the rms density contrast at $z$ over a sphere of comoving radius $r_M$ containing mass $M(r_M)$ as $\sigma_M(z)=[\frac{1}{2\pi^2}\int P_{\rm tot}(k,z)W_{\rm TH}(kr_M) k^2 dk]^{1/2}$ after normalizing to $\sigma_8$ over $8h^{-1}$Mpc at $z=0$. ($W_{\rm TH}$ is the top-hat function). Assuming spherical collapse, masses with density contrast $>\delta_{\rm col}=1.68$ at that epoch will have collapsed by $z$. In general, collapse and subsequent formation of compact objects is driven by balance between pressure and gravity, which in turn is determined by cooling in the collapsing gas. Two modes of halo collapse are  relevant here: if enough H$_2$ forms the gas will have $T\simeq 10^3$K and in the absence of metals the H cooling will in any event keep the gas at $T\simeq 10^4$K \citep[see review by][]{Bromm:2004}. Fig. \ref{fig:fig2} shows the resultant rms density fluctuation vs mass at various $z$ relevant here for $M_{\rm PBH}=(0, 15, 30) M_\odot$ with the vertical lines demarcating where the halo virial temperatures exceed these limits. The strong increase in the rms density contrast, over that in the absence of the PBHs, at masses of the first halos  capable of producing luminous objects, is obvious. This increase will lead to substantially more collapsed halos capable for forming luminous sources at $z>10$.

%1. LIGO measurements: equal BH mass, parameters
%
%2. Extra power
%
%3. $\sigma_M$ with the extra power

\begin{figure}[h!]
%\hspace{-13mm}
\includegraphics[width=7in]{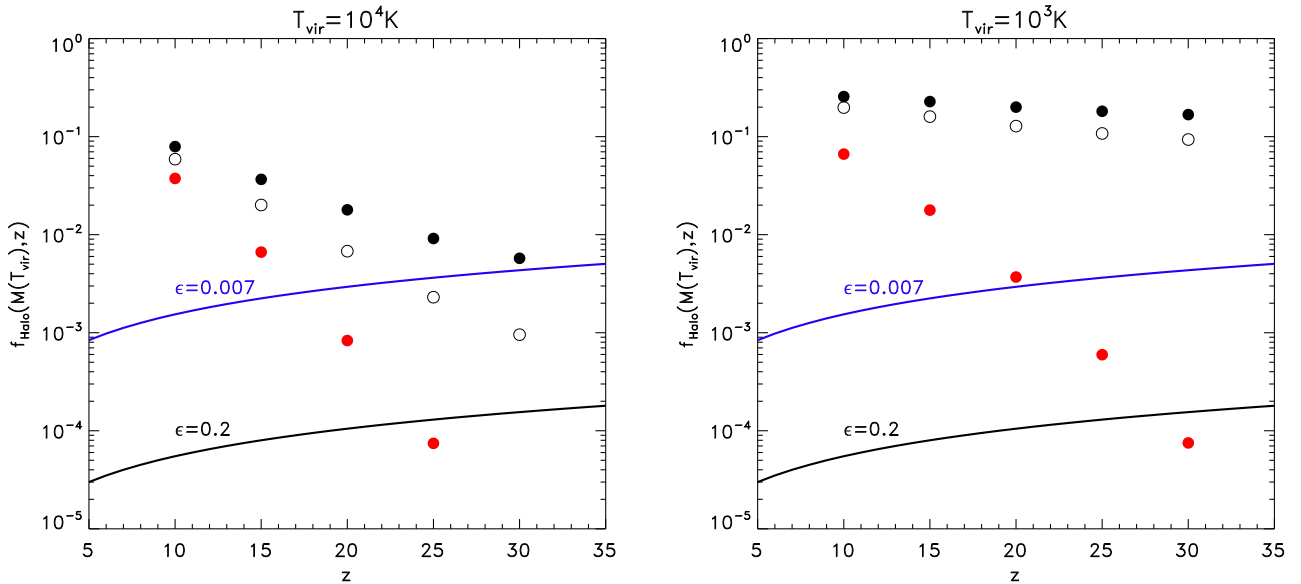}
 \caption[]{\footnotesize{Fraction of collapsed halos (eq. \ref{eq:f_halo}) at $T_{\rm vir}>10^4$K (left) and $T_{\rm vir}>10^3$K (right) vs $z$ for standard $\Lambda$CDM power spectrum (red filled circles), DM PBHs with $M_{\rm PBH}=15M_\odot$ (open black circles) and $M_{\rm PBH}=30M_\odot$ (filled black circles). Thick solid curves mark the overall fraction of baryons (effectively $f_*f_{\rm Halo}$) needed to produce the observed CIB per eq. \ref{eq:fraction_bc} with $f_{\rm Halo}=1$ with the H-burning radiation efficiency  $\epsilon=0.007$ (blue) and BH-type efficiency $\epsilon=0.2$ (black). The mean efficiency of the required conversion of baryons into luminous sources inside each halo would be the ratio of the solid curves to the circles. While $f_*$ is high (even higher than, or comparable to, 100\% at $z\gsim 20$), it remains very modest if the PBHs make up the DM.}}
\label{fig:fig3}
\end{figure}
We use  the Press-Schechter formalism \citep[][]{Press:1974} to compute the fraction of collapsed halos as the probability of a density field region with virial temperature $T_{\rm vir}$ having overdensity $>\delta_{\rm col}$. For Gaussian-distributed density fluctuation the fraction of the halos that collapsed by redshift $z$ is 
\begin{equation}
f_{\rm Halo} (M(T_{\rm vir}),z)=\frac{1}{2} {\rm erfc}\left(\frac{\delta_{\rm col}}{\sqrt{2}\sigma_{M(T_{\rm vir})}(z)}\right)
\label{eq:f_halo}
\end{equation}
Fig. \ref{fig:fig3} shows the fraction in halos that collapsed by $z$ having $T_{\rm vir}>10^4$ (right) and $10^3$K for $M_{\rm PBH}=(0, 15, 30) M_\odot$. The increase in $f_{\rm Halo}$ is large enough to produce the required CIB flux with very modest baryon conversion efficiencies (eq. \ref{eq:fraction_bc}) of well below $\sim 1\%$ for the H-burning $\epsilon=0.007$ even by $z\sim 20$ in halos with $H_2$-cooling. Even in halos with $T_{\rm vir}>10^4$K, the required $f_*$ remains at a modest few percent level at $z\simeq 12-15$ for $M_{\rm PBH}=30M_\odot$. If the bulk of the CIB comes from BH accretion, the values of the required $f_*$ drop by over an order of magnitude. Thus to account for the observed near-IR CIB fluctuation signal with high-$z$ emissions, very few baryons would need to be converted into luminous sources inside first collapsed halos at $z>10-15$ if the DM is made of PBHs such as discovered by LIGO.

\section{Discussion}
\label{sec:discussion}

If PBHs make up DM, luminous sources within the much more abundant early collapsed halos would reproduce the observed {\it Spitzer/Akari} source-subtracted CIB fluctuations with modest formation efficiency requirements. This can be demonstrated by taking population models from \cite{Helgason:2016} and rescaling them by the collapse-efficiency ratio from Fig. \ref{fig:fig3}. Thus Fig. 2, upper right from \cite{Kashlinsky:2015} would now reproduce the observed CIB signal with only $f_*<0.5\%$ forming out to $z\gsim15$ (instead of 4\% with formation continuing to $z\simeq10$) and the lines in Fig. 5 of \cite{Helgason:2016} need to be rescaled down by the corresponding factors. Additionally the measured CIB-CXB coherence \citep{Cappelluti:2013} would require that at least $\gsim$(10--15)\% of the luminous CIB-producing sources are accreting BHs, broadly consistent with this discussion. 

We now outline briefly the possible modifications in the early collapse and source formation that the PBHs may require. Two temperature regimes are relevant for description of emitting sources: 1) minihalos where H$_2$ formation is efficient evolve at $T\lsim 10^3$K and the gas converges toward density of $n_{\rm gas}\sim 10^4$cm$^{-3}$ \citep[][and refs therein]{Bromm:2004}, 2) in the absence of H$_2$, the metal-free gas will be able to cool to $10^4$K and collapse in halos with larger virial temperature will proceed isothermally. Feedback effects from first sources would affect H$_2$ formation via a resulting Lyman-Werner (LW) radiation at [11.2-13.6]eV \citep[see review by][]{Bromm:2013}. Gas collapse/evolution in the PBH minihalos may affect the subsequent emitting source formation inside them.

PBHs will accrete the minihalo gas resulting in both the additional source of emission from PBH accretion as well as the LW radiation feedback. The gas at sound speed $c_s$ within the halo of velocity dispersion $v_d$ will be accreted within the typical radius $r_{\rm acc}=GM_{\rm PBH}/u^2$ with $u^2=v_d^2+c_s^2$. The total accretion mass will be $M_{\rm acc} =2 (n_{\rm gas}/10^4 {\rm cm}^{-3}) (M_{\rm PBH}/30M_\odot)^3 (u/1{\rm km\;sec}^{-1})^{-6} M_\odot$. For typical parameters this may be a non-negligible fraction of the minihalo baryons at $\sim M_{\rm acc}/M_{\rm PBH}\times \Omega_{\rm CDM}/\Omega_{\rm bar}\propto M_{\rm PBH}^{2} u^{-6}$ up to a few percent, but will not increase the PBH mass dramatically. (Note the sensitive dependence on $u$, so $M_{\rm acc}$ is rapidly decreased when $u\gg1$km/sec). The spectrum of the resultant emission may be modeled after \cite{Yue:2013a}: 1) the multicolor black-body from different parts of the accretion disc with temperatures up to $T_{\rm max} \simeq 0.4 (M_{\rm PBH}/30M_\odot)^{-1/4}$keV shifted mainly into observer's near-IR, after reprocessing by the surrounding medium, and 2) hot corona and reflection emissions, which leave their mark in the observer soft X-rays. The coherence between the near-IR and X-ray emissions would be strong for the PBHs because of the larger value of $T_{\rm max}$ than for DCBHs. This mode of evolution, inevitable if PBHs make up DM, may influence adjacent star formation and DCBH collapse and evolution as discussed e.g. in \cite{Bromm:2003,Agarwal:2012,Yue:2014}.

The PBHs in minihalos will evolve via stellar dynamical effects similar to that discussed in \cite{Kashlinsky:1983} and by loss of energy to GW emissions \citep{Bird:2016}.  Stellar evaporation will lead to a core-halo structure with the isothermal core of radius $r_{\rm c}$ and $N_{\rm PBH}$ PBHs evolving on Gyr-timescales $t_{\rm evap}\sim N_{\rm PBH}/\ln N_{\rm PBH} \times r_{\rm c}/v_d$, at constant binding energy, or $v_d\propto N_{\rm PBH}^{-1/2}$, because evaporating PBHs carry zero energy. At the same time, a fraction of PBHs will become binary when GW emission exceeds their kinetic energy ($\sim v_d^2$); the cross-section for this process being $\sigma_{\rm GW} \simeq 10^{-8} (M_{\rm PBH}/30M_\odot)^2 (v_d/1{\rm km\; sec}^{-1})^{-18/7}{\rm pc}^2$ \citep{Bird:2016}. The fraction of PBHs that will form binaries before evaporation is then 
\begin{equation}
f_{\rm PBH, binary} \sim \frac{N_{\rm PBH}^2}{\ln N_{\rm PBH}}\frac{10^{-8}{\rm pc}^2}{r_c^2} \left(\frac{M_{\rm PBH}}{30M_\odot}\right)^{-2} \left(\frac{v_d}{1{\rm km\; sec}^{-1}}\right)^{-18/7}
\label{eq:pbh_binary}
\end{equation}
Instead of evaporating the resultant binaries will spiral in to the center due to dynamical friction possibly forming a central large BH contributing to the massive BH formation in early Universe.

Finally, we note further constraints from reionization by both first stars and BHs as discussed in this context in \cite{Atrio-Barandela:2014,Helgason:2016}. While reionization would be complicated in the presence of X-rays \citep{Ricotti:2004,Ricotti:2005}, the Thomson optical depth of $\tau\lsim 0.1$ may imply that the ionizing photons at rest $<0.0912\mic$ are mostly absorbed in their paternal minihalos, although \cite{Atrio-Barandela:2014} recover $\tau \lsim 0.05$ from hot gaseous bubbles reionized by first stars producing the observed CIB levels and forming to $z=9$. The situation, while important, is clearly model-dependent: emissions probed at observer $\lambda\gsim 2 \mic$, where current CIB data appear established, would translate directly into ionizing, Lyman-continuum, photons only at $1+z_{\rm ion} \geq 22(\frac{\lambda}{2\mic})$; at these epochs the baryon density is high and harder to reionize and only a small fraction of the CIB is expected to be produced. A possibility, discussed in \cite{Yue:2013a} for the DCBH model, whereby the gaseous collapsed halos are Compton thick so the ionizing photons are absorbed and reprocessed into a two-photon continuum, 
may also appear relevant here. %TBD: ionizing photons per 1 nW/m$^2$/sr as function of $z$ and $\lambda$.

%\begin{itemize}
%\item {\bf Accretion onto PBHs} DM halos, gas, etc. PBHs vs DCBHs \citep{Yue:2013a,Yue:2014}.
%\item {\bf Feedback effects and star formation in the PBH halos} H$_2$ formation etc \citep{Bromm:2004}, feedbacks \citep{Bromm:2013}. Star form when gas becomes self-gravitating at inner radii \citep{Kashlinsky:1982}. Also discuss IGM and cross-power w CMB \citep{Atrio-Barandela:2014}
%\item {\bf Stellar dynamical evolution of firs PBH halos} bla-bla  \citep{kashlinsky:1983}
%\item {\bf Effects of GW production on PBH growth}
%\item {\bf Reionization constraints and effects on IGM} 
%\end{itemize}

This work was supported by NASA/12-EUCLID11-0003 ``LIBRAE: Looking at Infrared Background Radiation Anisotropies with Euclid" project (\url{http://librae.ssaihq.com}). I thank my LIBRAE colleagues for comments on the draft of this paper.
\bibliographystyle{aasjournal}
%\bibliography{references_kashlinsky}

\end{document}